# Contribution of soundscape appropriateness to soundscape quality assessment in space: a mediating variable affecting acoustic comfort

Xinhao Yang [a,b], Guangyu Zhang [a,b], Xiaodong Lu [c], Yuan Zhang [a,b*], Jian Kang [d]

[a] Liaoning Provincial Key Laboratory of Eco-Building Physics Technology and Evaluation, Shenyang Jianzhu University, 25 Hunnan Middle Road, Shenyang, 110168, China
[b] School of Architecture and Urban Planning, Shenyang Jianzhu University, 25 Hunnan Middle Road, Shenyang, 110168, China
[c] School of Architecture and Fine Art, Dalian University of Technology, Dalian 116023, China
[d] UCL Institute for Environmental Design and Engineering, The Bartlett, University College London, 14 Upper Woburn Pl, London, WC1H 0NN, United Kingdom

[*] Corresponding author at: 25 Hunnan Middle Road, Shenyang, 110168, China.
 E-mail address: y.zhang@sjzu.edu.cn (Y. Zhang).

## Highlights

1. *SA* mediates the effect of sound source type on *AC*, and this relationship is spatially heterogeneous.

2. Ignoring the inhibitory effect of *SA* will lead to a misjudgement of *AC* effects by sound source type.

3. Spatial information can improve the precision of the interactions between variables.

## Graphical Abstract

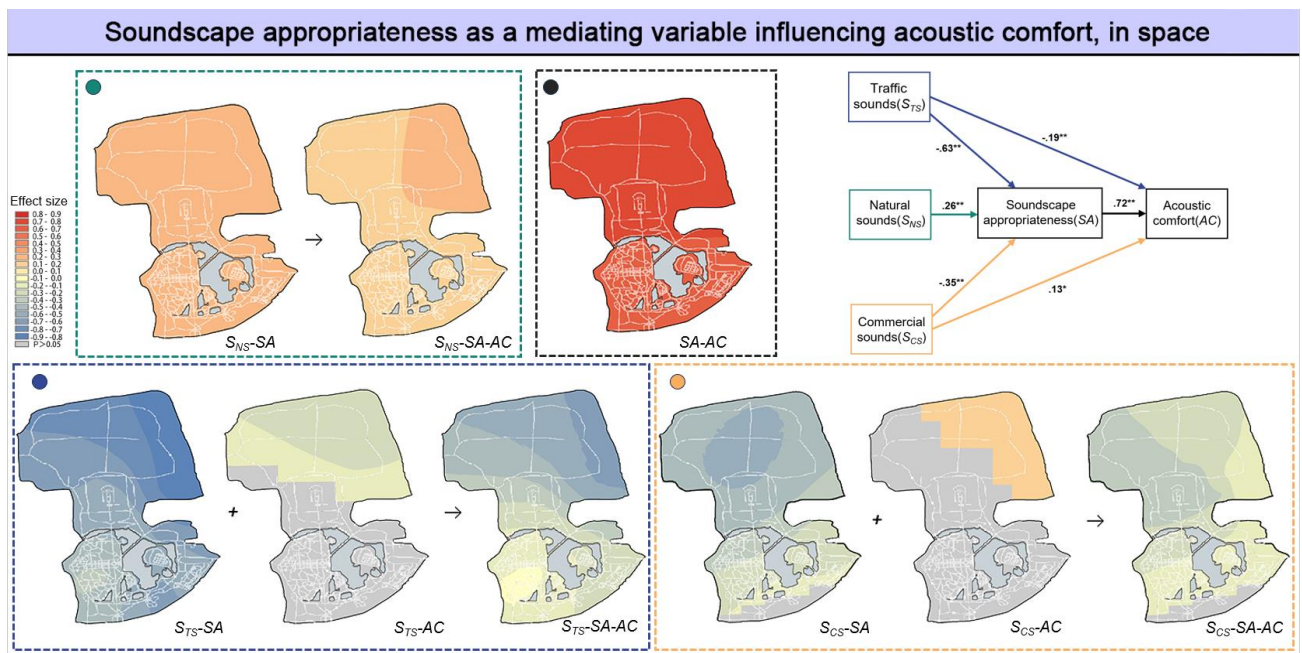

**Abstract:**

Soundscape appropriateness (*SA*) provides supplemental information on the matching degree between auditory information and the surrounding scene in soundscape perception. This indicator has been integrated into the standard ISO process for collecting soundscape data, forming a component of the sound quality assessment questionnaire. However, its role in soundscape quality assessment has not been fully understood. Herein, we present the findings from soundscape data collected from Beiling Park in Shenyang, China. A method was developed that integrates mediation effect models with multiscale geographically weighted regression models to explore the mediating role of SA in the impact of sound source types on soundscape quality, as well as the spatial heterogeneity of this mediation effect. The results confirm that SA does mediates the influence of sound source types on acoustics comfort (*AC*). Specifically, natural sounds (indirect effect/total effect = .19/.19), traffic sounds (indirect effect/total effect = -.46/-.65), and commercial sounds (indirect effect/total effect = -.25/-.12) impact the perception of AC by either enhancing or reducing SA. Moreover, the relationships among variables depicted in this model demonstrate spatial heterogeneity, demonstrating that in urban open spaces with complex constructures, local spatial models may be needed for soundscape assessment. The research reaffirms the significance of SA in urban open spaces. In terms of practical implications for urban and landscape planners, when sound sources cannot be controlled or altered, coordinating between the sound and the surrounding environment through landscape optimisation could also improve the quality of the soundscape through enhancing SA and help achieve the goal of creating healthy urban open spaces.

# 1. Introduction

The European Union's directive on the Assessment and Management of Environmental Noise has triggered a series of global actions aimed at diminishing noise levels (EUROPEAN UNION, 2002). Nevertheless, reducing sound levels does not equate to an enhancement in quality of life. In urban public open spaces where sound levels fall below certain thresholds, the connection between sound pressure levels ($SPL$) and people's acoustic comfort ($AC$) evaluation is not well related (Yang and Kang, 2005; Kang, 2007a). This is particularly highlighted in research during the COVID-19 lockdown. Despite the significant reduction in overall urban noise levels due to the blockade (Aletta et al., 2020), complaints concerning building and community noise have markedly increased. This growth is attributed to these noises becoming more noticeable and causing annoyance. (Tong et al., 2021).

Soundscape is defined as "acoustic environment as perceived or experienced and/or understood by a person or people, in context"(ISO, 2014). This definition establishes the concept of soundscape, which emphasises human experience and the perception of the surrounding sound environment. Past studies have confirmed the positive effects of high-quality soundscape on human well-being, physical and mental health, and quality of living environment (Zhang et al., 2015; Francis et al., 2017; Zhang et al., 2017; Aletta et al., 2018a; Jia et al., 2020; Buxton et al., 2021; Hong et al., 2022; Zhang et al., 2024). With further research, researchers are working to explore the factors that influence soundscape quality to improve the soundscape of a place.

After realising that sound can be seen as a resource rather than waste, researchers have dedicated themselves to exploring the emotional experiences that acoustic environments bring to people (Aletta et al., 2018b; Ratcliffe, 2021). In nondirecting environments, individuals' physiological and psychological systems selectively focus on specific sounds as auditory objects (Botteldooren et al., 2016), which then become the primary stimuli for soundscape perception. The origin of these sounds, referred to as sound sources, and the information they contain play a crucial role in the perception of the soundscape (Aletta et al., 2016; Axelsson et al., 2010; Li et al., 2024). Through the listener's perception of the sound sources, changes in response, emotion, and behaviour are generated, which ultimately leads to a perceptual result (ISO, 2014). Sound sources are usually classified as negative, positive or neutral according to the emotional experience they bring to people. For example, natural sounds, such as birdsong (Zhang et al., 2019) and water (Jeon et al., 2010; Hong and Jeon, 2015), are considered to have a positive impact on soundscape quality. The sounds of people talking and children playing are considered neutral in terms of AC (Dubois et al., 2006), while traffic (X. Zhang et al., 2018; Yang et al., 2019) and construction (Yang et al., 2021) sounds are considered negative.

However, the contribution of sound sources to the quality of a soundscape varies according to changes in the function of the locality (Hong and Jeon, 2015). In the perceptual structure of a soundscape, the understanding of the soundscape is highly dependent on context (Brown et al., 2016). The context includes the interrelationships between person and activity and place, in space and time (ISO, 2014). Previous studies have explored the relationship between context and soundscape perception, revealing that soundscape perception is influenced by the functionality of the environment and the physical attributes of the surroundings. (Pheasant et al., 2008; Watts et al., 2013; Jo and Jeon, 2022; Lu et al., 2022) Furthermore, researchers have found that people's perceptions of soundscape can be influenced by their expectations (Bruce and Davies, 2014), depending on their activities and behavioural intentions (Bild et al., 2018). When individuals enter a specific scene, they anticipate that the place will have a certain auditory environment inside (Kang, 2007b). When the actual sound environment conflicts with expectations, the individual's behaviour and evaluation of the soundscape and scene will be affected (Bruce and Davies, 2014). For example, in urban parks, inconsistencies between expected and actual soundscapes increase individuals' levels of annoyance (Brambilla and Maffei, 2006). In this case, assessing whether the soundscape is suitable for the surrounding environment will affect people's evaluation of the soundscape (Davies et al., 2014).

Soundscape appropriateness (*SA*) was proposed as an indicator of whether a soundscape is suitable for a place (Brown et al., 2011). Such appropriateness has considerable importance in terms of the consistency between soundscapes and environments. For example, in a church, enhancing the consistency between sound and the environment can increase the sense of presence (Larsson et al., 2007). The sound of human activities in the leisure areas of urban parks is considered to enhance spatial vitality, while traffic noise in green spaces in urban parks can increase visitors' annoyance (Bild et al., 2018). Many studies have examined the relationship between visual, auditory, and activity factors and SA through a variety of urban environments (Axelsson, 2015; Hamid et al., 2023; Hong et al., 2020; Jo and Jeon, 2020; Steele et al., 2015). A study that treated *SA* as an independent variable in the models and found a positive contribution to pleasantness (Cynthia et al., 2019). Obviously, the above evidence indicates the importance of *SA* for soundscape assessment (Axelsson, 2015; Jo and Jeon, 2020). Axelsson raised a question in the study on soundscape assessment (Axelsson, 2015): should appropriateness be the primary consideration in sound quality assessment? Although it is currently not possible to provide a definite answer to this question, delving into the role of *SA* in soundscape evaluation helps deepen our understanding of how people perceive soundscapes.

In the indicators of soundscape quality assessment, acoustic comfort (*AC*) represents the fundamental perception of users towards the sound environment (Z. Zhang et al., 2018). This indicator is widely used in relevant research on urban open spaces (Yang and Kang, 2005), scenic areas (Liu et al., 2024), and indoor environments (Gramez and Boubenider, 2017). Existing studies have confirmed that assessing *AC* can better represent park users' preferences for staying in urban parks (Tse et al., 2012). In prior studies, researchers have concentrated on and validated the influence of acoustic factors such as sound sources (Jeon et al., 2010) and environmental factors (Tsai and Lai, 2001) on *AC*.

Based on the evidence presented, SA may have played a role in the influence of sound source types on *AC*. In this context, sound sources in spaces have altered the *SA*, thereby affecting *AC*. In addition, the correlation among soundscape variables mirrors the attributes of spatial heterogeneity. This offers empirical support for further investigation into the spatial relationships between sound source types, *SA*, and *AC*. Through geographically weighted regression modelling, researchers have confirmed the spatial heterogeneity in the relationships between traffic noise, water sounds, bird calls, and soundscape quality at the urban scale (Hong and Jeon, 2017). In the study of scenic areas, researchers have identified the spatial heterogeneity of the relationships among sound source types, landscape satisfaction, audio-visual experience satisfaction, and perceived emotional quality (pleasure and eventfulness) with the sound environment (Chen et al., 2022). Based on this, the study established two specific research objectives.

1). Establish an influence path of "sound source type - *SA* – *AC*" and create a mediation effect model to analyse the mediating effect of SA on the relationship between sound source type and *AC*.
2). Establish a multiscale geographically weighted regression (*MGWR*) model to explore the spatial heterogeneity of the mediating effect.

## 2. Methods

*2.1 Model structure*

The integrated model approach used here combines the mediation effect model method and the *MGWR* model method (Fig. 1). A mediation effect model identifies and tries to understand potential mediating variables that serve as intermediaries between the independent and the dependent variables (Wen and Ye, 2014). Specifically, the mediation effect model postulates that the independent variable $x$ influences the dependent variable $y$ through a mediating variable $M$ (Preacher and Hayes, 2008).This process can be separated into two primary paths: first, the effect of the independent variable on the mediating variable; and second, the effect of the mediating variable on the dependent variable. When both paths are simultaneously present, mediation effects become apparent, indicating that the relationship between the independent and dependent variables occurs in part indirectly through the mediating variable.

In this study, we designated *SA* as the mediating variable, with sound source type and *AC* serving as the independent and dependent variables, respectively, within the mediation effect model (see Fig. 1). This model can reveal the direct effect of sound source type on *AC*, as well as the indirect effect of sound source type on *AC* through *SA*. Moreover, after integrating *MGWR*, this model can reflect in spatial resolution the above relationships; that is, as spatial location changes, the mediating effect of *SA* changes.

Assuming that there are $n$ observed values and that the observed value $i$ belongs to locations $u_i$ and $v_i$, the model can be expressed as follows:

$$y_i = \sum_{j=0}^{M} \gamma_{b\omega j}(u_i, v_i) x_{ij} + \varepsilon_{1i}$$

$$M_i = \sum_{j=0}^{M} \alpha_{b\omega j}(u_i, v_i) x_{ij} + \varepsilon_{2i}$$

$$y_i = \sum_{j=0}^{M} \gamma'_{b\omega j}(u_i, v_i) x_{ij} + \beta_{b\omega j} M + \varepsilon_{3i}$$

where $x_{ij}$ is the $j_{th}$ predictor variable, $\gamma_{b\omega j}(u_i, v_i)$ is the $j_{th}$ coefficient, $\varepsilon_i$ is the error term, $y_i$ is the response variable, and $M_i$ is the mediator variable. $b\omega j$ in $\gamma_{b\omega j}$ indicates the bandwidth used for calibration of the jth conditional relationship. The same notation is used for $\alpha$ and $\beta$.

*2.2 Study area*

Beiling Park was built on the Zhaoling Mausoleum, the ancient emperor's tomb, and its surrounding green spaces (Fig. 2). It is the largest urban park in the local area, covering an area of 3.03 million square meters, of which the water area is 230,000 square meters. Marked by the yellow dotted line in the diagram, the park is divided into two parts: a northern part and a southern part. To the north of the park is the Mausoleum area of Emperor Taiji of the Qing Dynasty (a world cultural heritage site) and an ancient pine forest area. These areas are also the main attractions for most nonlocal tourists. A modern urban park area was built in the southern part. This area contains rich urban green spaces, water bodies, leisure places, and sports and fitness places, which attract a large number of local citizens. Therefore, a large soundscape contrasts with the peaceful and solemn environment of the mausoleum area in the park. Moreover, Beiling Park is located in a high-density city surrounded by urban roads on all sides, and traffic from nearby urban areas interferes with the park. In such an urban park with multiple functions and intertwined sound sources, it is worth examining whether the soundscape is appropriate for the surrounding landscape environment.

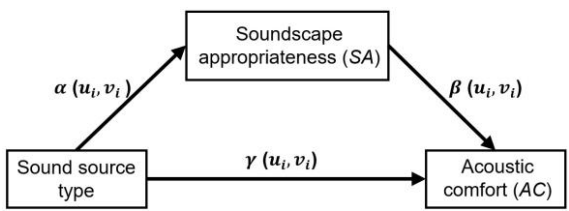

**Fig. 1** Theoretical relations of the integrated model

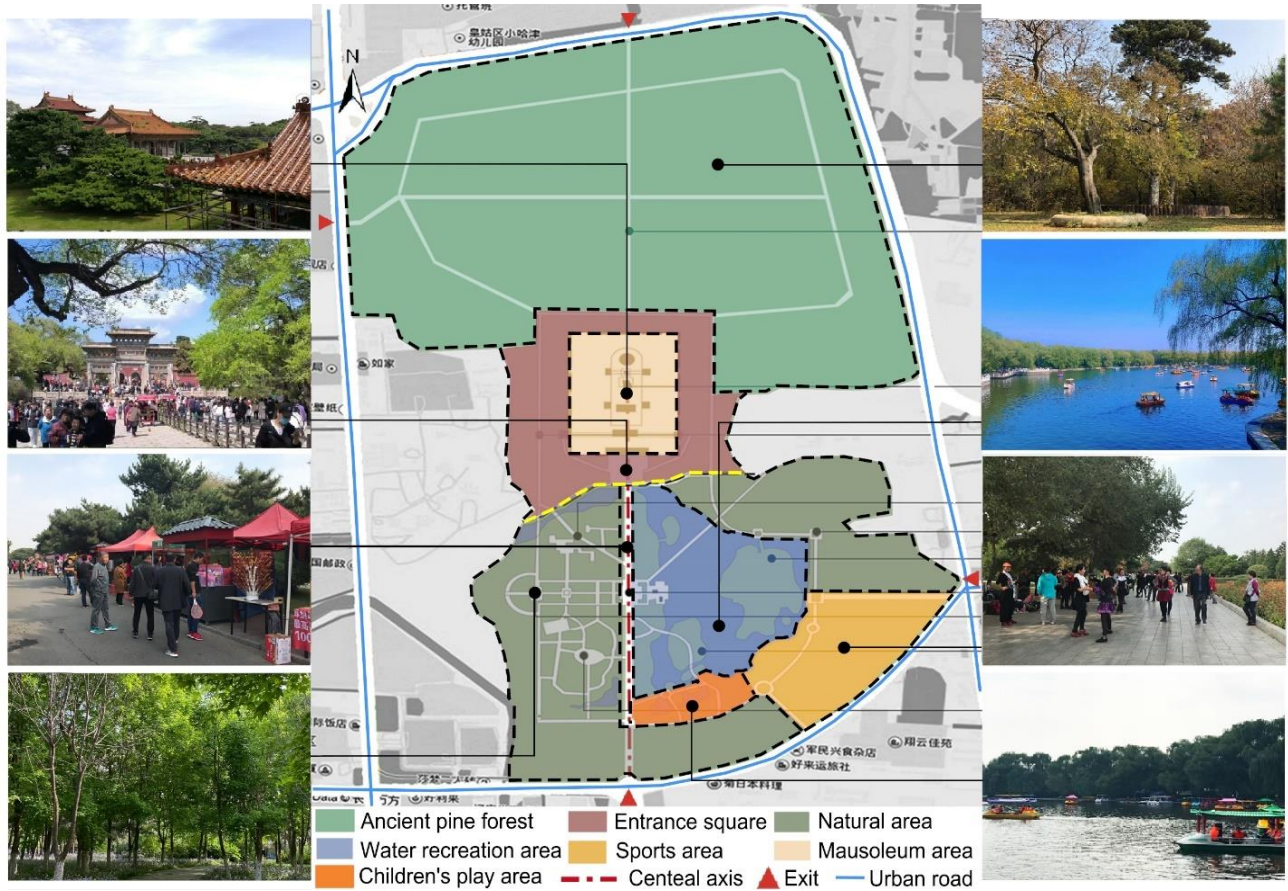

**Fig. 2** Functional division of Beiling Park.

*2.3 Data collection*

The soundscape assessment was performed according to the guidelines of the European Research Council Soundscape Indices (SSID) project (Aletta et al., 2020; Mitchell et al., 2020) (Table 1). The assessment was conducted on a single day of a weekend in the autumn (from 10 am to 2 pm). Visitor visitation and recreational activities were abundant and stable during this period. The weather conditions during the measurements were stable, with no rain, snow, thunder or lightning. The soundscape features remained relatively stable throughout the measurement period, ensuring the synchronicity of the soundscape data.

The data collection was divided into two parts: the *SPL* measurement and the soundwalk. For the *SPL* measurements, 205 measuring points were identified based on the distribution of attractions and tourists in the park. Sixteen surveyors used a Circus CR: 162B sound level meter to measure $L_{Aeq}$ for 30 s at the measuring points. For the soundwalk, the park was redivided into 128 grids (the grid size ranges from 100m to 400m.) according to spatial characteristics and the landscape and soundscape composition (*Fig. A1*). Eight assessors worked in groups of two to assess the soundscape of the park through a soundwalk.

The sound source saliencies, *SA*, and *AC* were evaluated by assessors, during the soundwalk. The sound source saliencies mean how significant or salient a particular sound source type is, serving as a measure of the perceptibility of it. This is determined by the energy of this sound source type and its prominence relative to other sound source type that exist simultaneously. Based on ISO and SSID project classifications of sound sources, we have identified five types of sound sources for Beiling Park based on preliminary investigations. ISO categorises sound sources in urban sound environments into three main types: sounds of technology, sounds of nature, and sounds of human activity (ISO, 2018). The SSID project studying urban open spaces specifically identified traffic noise as a separate source type, along with other noises, sounds from human activity, and natural sounds (Mitchell et al., 2020). In this study, the typical sounds present in Beiling Park are categorised into natural sounds ($S_{NS}$), traffic sounds ($S_{TS}$), commercial sounds ($S_{CS}$, e.g., the calls of peddler), recreational sounds ($S_{RS}$, e.g., music played by amusement facilities and laughter from crowds), and tourist sounds ($S_{ToS}$, e.g., conversations and footsteps of tourists during their visits).

**Table 1**
Soundscape data collection information.

| | Indictors | Possible response |
|---|---|---|
| Sound type saliency | Natural sounds ($S_{NS}$) Traffic sounds ($S_{TS}$) Commercial sounds ($S_{CS}$) Recreational sounds ($S_{RS}$) Tourist sounds ($S_{ToS}$) | 1: Do not hear at all - 5: Dominates completely |
| Descriptors of soundscape | Soundscape appropriateness (*SA*) | 1: Not appropriate at all - 5: Definitely appropriate |
| | Acoustic comfort (*AC*) | 1: Not at all - 5: Definitely comfortable |
| Sound pressure level, SPL ($L_{Aeq}$, 30s) | | |

To avoid possible order effects, the four groups of evaluators followed different paths during the sound walk (Fig. A1). Each group started from the southern/northern entrance of the park and completed the circular sound walk in the clockwise and counterclockwise directions, respectively (ISO, 2018). At the centre of each grid point, the sound source saliencies, *SA*, and *AC* were evaluated using a 5-point Likert scale, and the latitude and longitude coordinates of the grid points were recorded. All assessors self-reported normal hearing and received training in soundscape assessment beforehand. Due to variations in the timing at which the four groups of personnel passed through the grid points, the collected soundscape data represent the average acoustic features during the measurement period.

*2.4 Statistical analysis*

*2.4.1 Mediating effect model*

Utilising linear regression to screen the variables included in the mediation effect model, linear regression models were established with *SA* and *AC* as dependent variables. Five sound source saliences were considered as independent variables. In the *AC* model, the independent variable also includes the hypothesised mediator variable *SA*. *SPL* was not included as an independent variable in the mediation model of sound source types as it is a comprehensive index determined by the composition and energy performance of various sound sources in space. This provides overlapping information with the intensity information of sound source saliencies.

**Table 2**
Linear regression results of *SA* and *AC*.

| DV | IV | β | P | VIF | F(P) | R² |
|---|---|---|---|---|---|---|
| *SA* | $S_{TS}$ | -.587 | .000 | 1.839 | 28.539 (.000) | .539 |
| | $S_{NS}$ | .271 | .002 | 1.957 | | |
| | $S_{CS}$ | -.364 | .001 | 1.521 | | |
| | $S_{RS}$ | -.074 | .505 | 3.202 | | |
| | $S_{ToS}$ | .156 | .112 | 2.501 | | |
| *AC* | $S_{TS}$ | -.169 | .032 | 2.586 | 50.703 (.000) | .715 |
| | $S_{NS}$ | .106 | .136 | 2.116 | | |
| | $S_{CS}$ | .132 | .046 | 1.808 | | |
| | $S_{RS}$ | .113 | .197 | 3.214 | | |
| | $S_{ToS}$ | -.068 | .385 | 2.553 | | |
| | *SA* | .701 | .000 | 2.170 | | |

The linear regression results for *SA* and *AC* are presented in Table 2. Following the exclusion of the insignificant $S_{ToS}$ and $S_{RS}$ variables, a mediation effect model was formulated with *SA* as the mediator variable and *AC* as the dependent variable. The model was subsequently validated using Amos 27 software. Four indicators are used to test the model: Comparative Fit Index (CFI), Root Mean Square Error of Approximation (RMSEA), Standardized Root Mean Square Residual (SRMR), and chi-square/degree of freedom (CMIN/DF).

*2.4.2 Spatial dependence of soundscape variables*

Spatial autocorrelation analysis was performed on the sound source saliency, *SA* and *AC* data to determine the correlation between data types in the same spatial grid and the same type of data in neighbouring grids. The spatial autocorrelation coefficient (Moran's I) was used to describe spatial autocorrelation. As shown in Table 3, within the study area, the sound source saliencies, *SA*, and *AC* all exhibited significant positive spatial autocorrelation (Moran's I > 0, | z | > 2.58, P value < .01). This indicates that the similarity of variables increases as the spatial distance decreases, showing the spatial dependence of the data.

**Table 3**
Spatial autocorrelation of the soundscape variables.

| Variables | $S_{NS}$ | $S_{TS}$ | $S_{CS}$ | $S_{RS}$ | $S_{ToS}$ | AC | SA |
|---|---|---|---|---|---|---|---|
| Moran's I | .41 | .29 | .44 | .50 | .31 | .24 | .19 |
| z | 14.11 | 10.05 | 14.94 | 18.62 | 10.78 | 8.18 | 6.53 |
| P value | .00 | .00 | .00 | .00 | .00 | .00 | .00 |

*2.4.3 Spatial heterogeneity analysis of the relationships among soundscape variables*

Based on the model relationships constructed in this study, the spatial heterogeneity between the significance of sound sources and *AC* when *SA* acts as an intermediary variable was explored using the MGWR method. Geographically weighted regression (GWR) is a method used to explore spatial heterogeneity relationships between variables, and it is widely used and applied to the fields of vegetation distribution (Kim and Shin, 2016), urban surface temperature (Song et al., 2014) and urban growth (Gao et al., 2020). This study confirmed that the classic geographic weighting regression model was also suitable for the noise (Ryu et al., 2014) and soundscape perception (Hong and Jeon, 2017) GWR models in urban environments and used a single kernel function and bandwidth parameter to calculate the weights (Li and Fotheringham, 2020). The bandwidth parameter controls the amount of distance attenuation when weighting the neighbours around each location. The optimal bandwidth in the GWR model reflects the spatial variation in all parameter estimates at the "best mean" scale of the spatial relationship. This also causes the spatial variations in all parameter estimates to exhibit the same scale characteristics.

Furthermore, the multiscale geographically weighted regression (MGWR) method was proposed to adapt to the relationships among multivariate spatial data corresponding to different change scales (Fotheringham et al., 2017). This method removes the single bandwidth assumption and allows the relationship between each independent variable and the dependent variable to vary according to different space-dependent scales.

## 3 Results

*3.1 Spatial distribution of the soundscape variables*

We used ArcGIS 10.8 to perform spatial interpolation on the soundscape data collected in this study and produced soundscape maps to observe the spatial distribution characteristics of the variables. Fig. 3(a-f) shows the spatial distribution characteristics of the *SPL* and sound source saliencies in the park. The overall *SPL* in the park ranged from 39.8 to 68.3 dB (A), and the high-SPL areas were mainly concentrated in the eastern and northern regions of the ancient pine forest area, both sides of the central axis, the children's play area, the sports area, and the area near the high-traffic roads in the southern region. By comparing the *SPL* distribution with the sound source saliency distribution, it can be found that the dominant sound sources in the high-SPL area are different. The areas near the external roads (the east and north sides of the ancient pine forest area and the south side of the park) were more affected by traffic noise (Fig. 3(c)). Commercial sounds were dominant in the children's play area and sports area (Fig. 3(e)) because there are many leisure and entertainment facilities there. Tourists are concentrated in this area and are mainly engaged in entertainment and leisure activities. The abovementioned areas were also accompanied by significant commercial activities; therefore, commercial sounds were also prominent (Fig. 3(f)). Natural sound was dominant in the low-SPL area (Fig. 3(b)), while the saliencies of other types of sound sources in the area with prominent natural sound were generally low. Most areas of the park were filled with the voices of tourists (Fig. 3(d)). However, these sounds are generated by walking and talking, and therefore, the sound level is low.

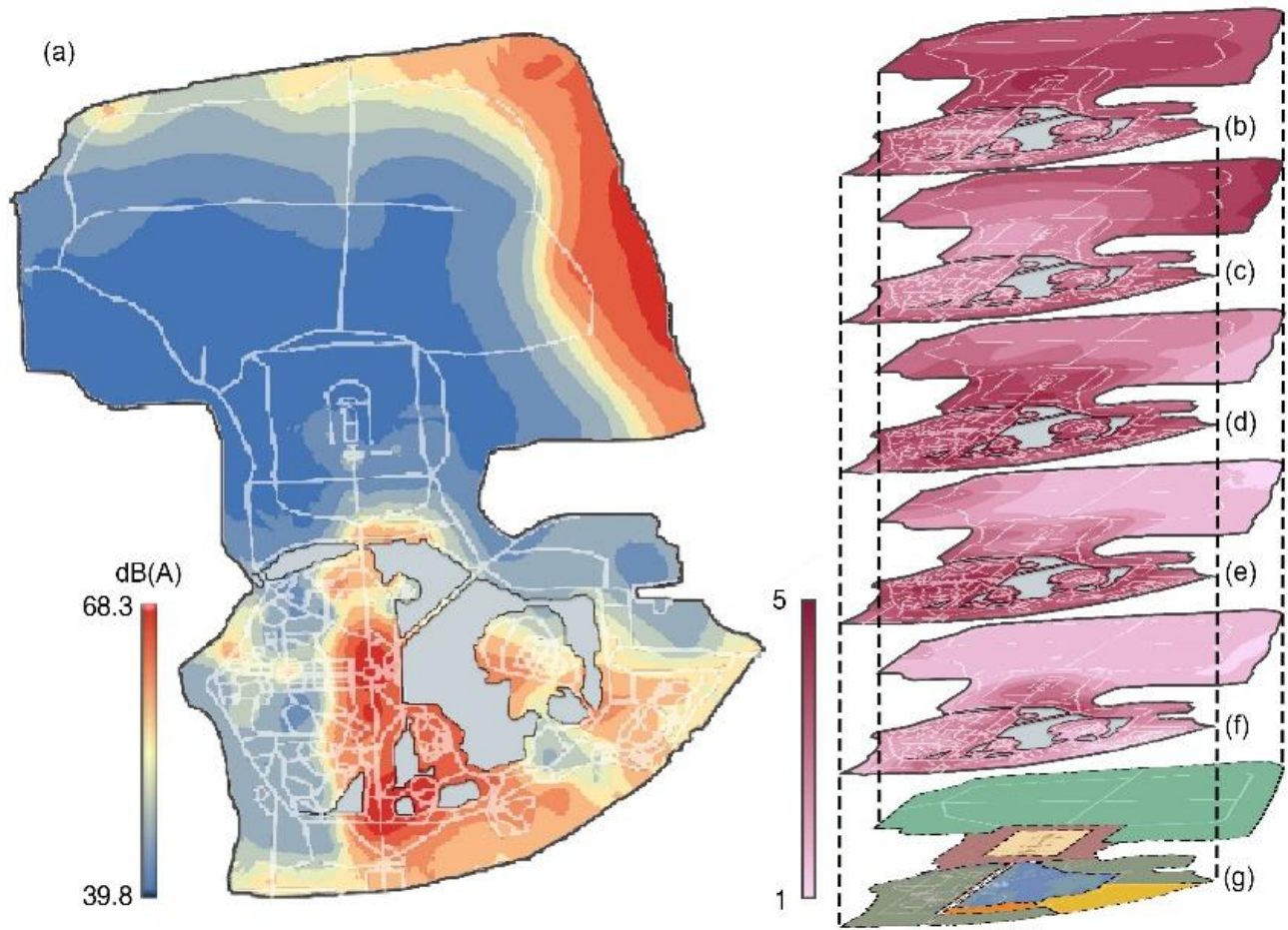

**Fig. 3** The spatial distribution of the *SPL* and sound type saliency: (a). *SPL*; (b). $S_{NS}$; (c). $S_{TS}$; (d). $S_{ToS}$; (e). $S_{RS}$; (f). $S_{CS}$; (g). Functional division (legend reference *Fig. 2*)

Fig. 4 shows the mapping results of *SA* and *AC*. The areas with higher *ACs* were mainly concentrated in the ancient pine forest area, the northern mausoleum area, and the natural area. Combined with the spatial distribution of the salience of sound sources, these findings indicate that these areas had more natural sounds and lower *SPL*. In contrast, the *AC* performance in areas with higher *SPL* was poorer. Similar to the distribution characteristics of *AC*, high *SA* was also distributed in areas where natural sound was salient. Furthermore, the *SA* and *AC* were lower in areas more influenced by traffic sounds and in other high-*SPL* areas.

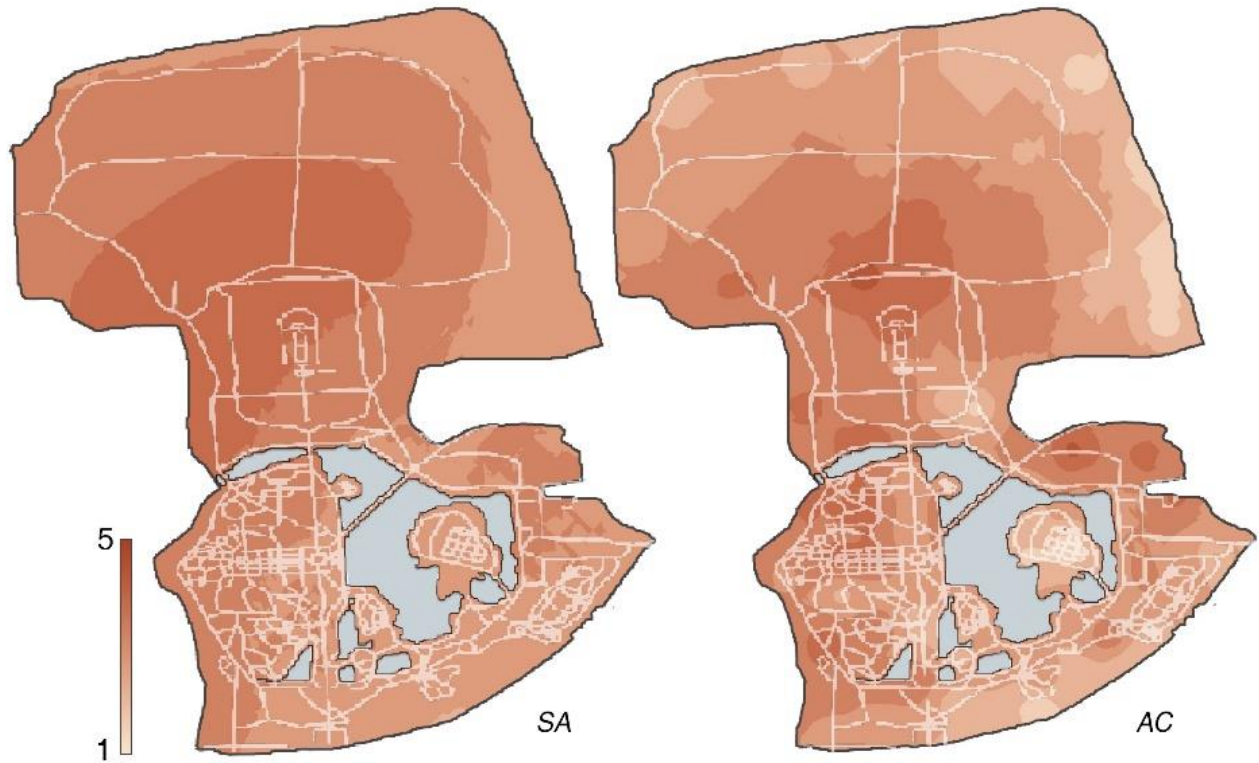

**Fig.** 4 The spatial distributions of *SA* and *AC*.

## 3.2 Global mediating effect of soundscape appropriateness

To investigate the potential mediating role of *SA* in the impact of sound source salience on *AC*, a mediation effect model was constructed with five types of sound source salience indicators as independent variables, *SA* as the mediating variable, and *AC* as the dependent variable.

This study concentrates on examining the influence of different types of sound sources on the soundscape. Despite the absence of multicollinearity, we maintain that incorporating the *SPL* in the model could alter the significance of sound sources, as the importance of sound sources inherently encompasses aspects of sound intensity. Consequently, we opted not to include *SPL* in the mediation model. After excluding nonsignificant types of sound source salience, $S_{ToS}$ and $T_{RS}$, an analysis of the mediating effect of *SA* was conducted.

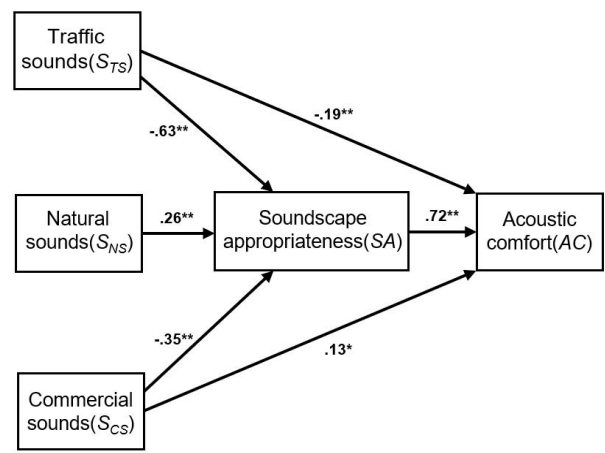

**Fig. 5** The path relationships in the mediation effect model.

The results of the mediating effect model showed that *SA* mediated the effect of the saliencies of the sound source on *AC*. Fig. 5 shows the path effect relationship under the mediating effect of *SA* in the mediation effect model.

**Table 4**

Results of the mediation effect.

| Path | | β | S.E | C.R. | P value | Direct effect | Indirect effect | Total effect |
|---|---|---|---|---|---|---|---|---|
| | SA→AC | .72 | .01 | 10.90 | .000 | | | |
| $S_{TS}$→SA→AC | $S_{TS}$→SA | -.63 | .05 | -9.07 | .000 | -.19 | -.46 | -.65 |
| | $S_{TS}$→AC | -.19 | .04 | -2.66 | .000 | | | |
| $S_{NS}$→SA→AC | $S_{NS}$→SA | .26 | .06 | -4.88 | .000 | | .19 | .19 |
| $S_{CS}$→SA→AC | $S_{CS}$→SA | -.35 | .05 | -4.88 | .000 | .13 | -.25 | -.12 |
| | $S_{CS}$→AC | .13 | .04 | 2.13 | .034 | | | |

The fit of the model was good, with a robust CFI = .998 (>.95) (Hu and Bentler, 1999), RMSEA = .063 (<.08) (Browon and Cudeck, 1993), SRMR=.013 (<.09) (Hu and Bentler, 1999), and CMIN/DF = 1.509 (<2) (Byrne et al., 1989). The models explained 53% (*SA*) and 71% (*AC*) of the variance. Table 4 lists the effect sizes for each path. $S_{TS}$ (total effect = -.65) and $S_{CS}$ (total effect = -.12) had negative effects on *AC* through *SA*, while $S_{NS}$ had a positive effect on *AC* through *SA* (total effect = .19). *SA* partially mediated the effects of $S_{TS}$ (indirect effect = -.46) and $S_{CS}$ (indirect effect = -.25) on *AC* and completely mediated the effect of $S_{NS}$ (indirect effect = .19) on *AC*.

*3.3 The spatial distribution of the impact of sound source salience on AC*

Using Mgwr 2.2 software, the MGWR models of *SA* and *AC* were created. The default Gaussian function is used as the kernel function to estimate the local coefficients and bandwidths. The adaptive kernel method is used to control bandwidth selection, and the golden section method is used to determine the optimal bandwidth size (Oshan et al., 2019). Table 5 provides relevant information on the MGWR model. The autocorrelation test of the model errors showed that the residuals of the two models did not exhibit spatial autocorrelation (|z| < 2.58, P value > .01). This indicates that the residuals are randomly distributed in space and that the model is reliable. In addition, the bandwidths for the variables in the MGWR model varies between 71 and 127 showing value of the multiscale nature of the employed model.

**Table 5**

Summary of model fit indices for the MGWR model.

| Model | Bandwidths | | | Adjusted R² | Aicc | Residual autocorrelation | | |
|---|---|---|---|---|---|---|---|---|
| | $S_{NS}$ | $S_{TS}$ | $S_{CS}$ | | | z | P value | Moran's I |
| SA | 126 | 71 | 88 | .57 | 270.93 | 1.28 | .20 | .03 |
| | SA | $S_{TS}$ | $S_{CS}$ | Adjusted R² | Aicc | z | P value | Moran's I |
| AC | 127 | 122 | 127 | .75 | 200.92 | .07 | .95 | -.06 |

Based on the model relationship in Fig. 5, the spatial distribution of the effect size for each path was obtained through interpolation, as shown in Fig. 6. The total effect size of each path was determined by combining the direct and indirect effect sizes ($S_{TS}$-SA-AC, $S_{NS}$-SA-AC, $S_{CS}$-SA-AC, in Fig 6). The purpose of this study was to explore the spatial distribution characteristics of the effects of sound source salience on *AC* under the mediating effect of *SA*. In the calculation process, the data that did not meet the significance requirement were excluded. Fig. 6(b) shows that $S_{NS}$ had a positive effect on *AC*, and the difference in the spatial distribution of effect sizes was not obvious. In contrast, $S_{TS}$ and $S_{CS}$ had negative effects on *AC* (Fig. 6(a) and (d)), and the difference in the effect sizes varied more spatially, with the difference in the effect size of traffic sounds being more significant. In the ancient pine forest area, the negative effect of $S_{TS}$ on *AC* was greater than that in the southern area. however, the negative effect of $S_{TS}$ on *AC* was relatively small but still present. In the mausoleum area and the western part of the ancient pine forest, the negative effect of $S_{CS}$ on *AC* was greater than that in other areas.

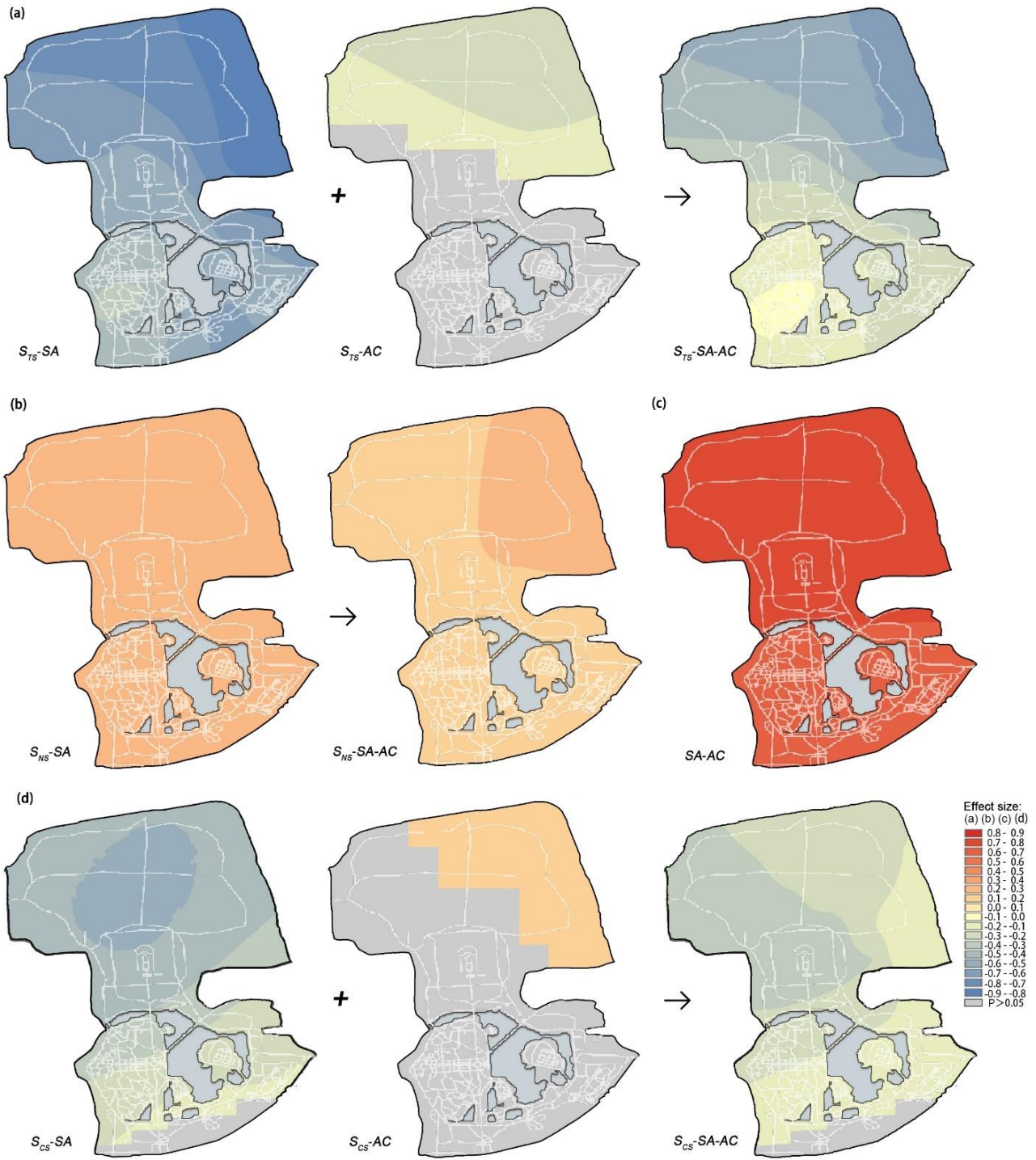

**Fig. 6** Spatial analysis results of the soundscape variable relationships: (a). The spatial distribution of impacts under the $S_{TS}$-*SA*-*AC* path; (b). The spatial distribution of impacts under the $S_{NS}$-*SA*-*AC* path; (c). The effect size of *SA* on *AC*, in space; (d). The spatial distribution of impacts under the $S_{CS}$-*SA*-*AC* path.

## 4 Discussions

*4.1 Role of soundscape appropriateness in soundscape assessment*

In this study, we theorised a model and conducted an analysis that included *SA* as a mediating variable in the effect of sound source saliencies on *AC*, and at the same time, this mediating effect was characterised by spatial heterogeneity. The assessment of *SA* can supplement the information provided by soundscape quality assessments (ISO, 2014). The study noted that this indicator should not be used alone. The assessment of *SA* cannot completely replace the assessment of soundscape quality (Axelsson, 2015). This study investigated the role of *SA* in the assessment of soundscape. By validating theoretical models, it is possible to explain the mediating effect of SA on the interaction of acoustic variables.

The results of this study indicate that part of the influence of sound sources on *AC* can be explained by the mediating role of *SA*. In Beiling Park, traffic noise disrupts the *SA* (indirect effect = -.46), leading to a negative impact on *AC* (total effect = -.65). Conversely, natural sounds enhance SA (indirect effect = .19), resulting in a positive influence on *AC* (total effect = .19). In addition, we observed an inhibitory effect of commercial sounds under the mediation of *SA* (Table 4). The direct effect of commercial sound on *AC* was positive (effect = .13), while the mediated effect was negative, with a larger absolute value (effect = -.25). This was specifically reflected in the spatial distribution of the effect size (Fig. 6(d)). In urban spaces, the sounds of human activities (nonmotorised, nonmechanical) can represent a manifestation of vitality (Zhao et al., 2021; Gan et al., 2022). The commercial sounds in Beiling Park can be seen as a source of sound that enhances spatial vitality in urban parks, which may be the reason why commercial sounds have a positive direct impact on *AC*. However, the presence of commercial sound destroyed the appropriateness of the soundscape in the area, thus having an inhibitory effect. The emergence of such sounds creates a conflict with the peaceful atmosphere cultivated in that area. This result suggests that ignoring such mediating effects may lead to erroneous statistical inferences.

Compared to previous studies which have primarily focused on the relationships between sound source and *SA* (Nielbo et al., 2013; Axelsson, 2015), or between *SA* and soundscape quality (Cynthia et al., 2019; Jo and Jeon, 2020). This study integrates these three elements, revealing the contribution of SA in the path of sound source affecting AC. This emphasises the role of SA in soundscape assessment and provides empirical support for a deeper understanding of soundscape perception processes. Meanwhile, the findings of the research can provide ideas for soundscape optimization from a design and management perspective. On the one hand, creating positive and appropriate sound sources, while eliminating inappropriate ones, can effectively enhance the quality of the acoustic environment. On the other hand, when sound sources cannot be controlled or altered, coordinating between the sound and the surrounding environment through landscape optimisation and environmental atmosphere creation is a reliable means to improve the quality of the soundscape and achieve the goal of creating healthy urban open spaces. Based on the mediating effect of *SA* on the effect of the sound source on *AC*, it is suggested that the appropriateness of the soundscape can be increased through the optimisation of the landscape environment and environmental atmosphere creation, which can then effectively enhance the soundscape. For example, the visual atmosphere or environmental ambience of commercial activities in areas with prominent commercial sound can be enhanced to increase the appropriateness between the sound source and the surrounding environment and enhance positive or reduce negative soundscape experience.

*4.2 Spatial heterogeneity of soundscape variables*

The spatial relationships between the soundscape variables explored in this study suggest that spatial regression methods are more effective for investigating local spatial relationships among soundscape variables, a finding that supports previous research (Hong and Jeon, 2017). Many studies use linear regression models to examine the relationships between soundscape variables (Axelsson et al., 2010; Axelsson, 2015; Rey Gozalo et al., 2015), which assumes that the relationships among these variables to be stable in space. Although this method can explain the general relationship between variables within the entire area, it may overlook the specific manifestations of the above relationships in specific localities brought about by the spatial distribution of features such as functions, landscapes, and facilities (Hong and Jeon, 2015). In this study, the *AC* in the northern area of Beiling Park, which is mainly composed of natural landscapes, was more affected by the negative impacts of traffic and commercial sounds than was that in the southern area of the park, which is mainly composed of man-made landscapes. This is mainly caused by the function and characteristics of the landscape space and by the mutual masking between sound sources. This finding also supports the results of previous studies that spatial factors may lead to non-stationarity in the spatial relationships of soundscape variables (Hong and Jeon, 2017).

Previous studies have used GWR to reveal the spatial variations of relationships between soundscape variables. (Hong and Jeon, 2017; Chen et al., 2022). The model constructed in this study uses the MGWR method, which allows variables to have their own bandwidths, meaning that it can adapt to different spatial variabilities of different variable. This made the model more accurate in capturing the spatial relationships between variables.

The many benefits that urban parks bring to the ecological environment and public communities have prompted people to have a keen interest in understanding how good urban park design can attract more tourists or prolong their stay. The analysis of the relationships among the soundscape variables in local areas in this study can provide valuable insights for soundscape optimisation and planning in urban parks. The spatial distributions of the effects of sound source saliencies on AC obtained from the spatial mediating effect model are helpful for determining the focus of soundscape optimisation in specific areas. This information is valuable to urban management departments, decision-makers, and landscape planners, and it helps to minimise unnecessary economic and resource waste through targeted and optimised design of the soundscape.

Specifically, in Beiling Park (Fig. 6(a)), in areas near urban roads (northeast, southeast, and south), the negative effect of traffic noise on soundscape comfort was significant. Therefore, special attention should be given to the prevention and control of traffic noise in this area. Even though the negative effect decreased deeper into the park, it still resulted in a decrease in *AC*. Therefore, the prevention and control of traffic noise still need to be strengthened overall. In the mausoleum area and its northwestern region, commercial sound had a significant negative impact on the *AC*. It was shown that there was a lack of soundscape transition between the bustle of the southern area of the park and the tranquillity of the tomb area (Fu et al., 2018). The *AC* of the tomb area was mainly affected by the sounds of the entertainment activities in the southern area and the commercial sound from the transition area between the two parts. Therefore, in this area, commercial activities should be reduced, or sound masking should be increased to maintain the original solemn atmosphere of the tomb site. In addition, in the overall area of Beiling Park, the amount of natural sound should be increased to improve the *AC* in the park. The effect of natural sound on *AC* was relatively homogeneous throughout the entire study area. This result is consistent with the results of a previous study conducted in tourist areas, which showed that geophysical sound had a consistent positive effect on the pleasantness dimension of soundscape perception (Chen et al., 2022). These results support the view that the positive effect of natural sounds on soundscape perception is stable.

*4.3 Limitations and future studies*

The models and methods developed in the present study can be used to guide soundscape investigations, assessments, and optimisations of urban parks and similar spaces. The research data was derived from urban parks that have been transformed from ancient imperial sites. The specific effects among the variables require further verification to assess their applicability to other types of city parks. Nevertheless, the methods and exploration of variable relationships in this study can serve as valuable references for related research.

Additionally, the data in this study is derived from a single time point and does not take temporal factors into account. Studies have shown that the soundscape changes over time (Liu et al., 2013; Chen et al., 2024). In the future, exploring the changes in soundscape relationships in the temporal dimension using geographical and temporal weighted regression (GTWR) (Fotheringham et al., 2015), Which is suitable for problems where relationships between variables vary across both space and time, will helps to provide a more comprehensive understanding of soundscape.

For empirical research on soundscapes, field data collection is the foundation and characteristic of soundscape empirical research because reliable primary data can be obtained (ISO, 2018). However, this requires a large amount of fieldwork and data processing. Currently, machine learning technology has been used in soundscape research for tasks such as urban street view (Zhao et al., 2023) and audio recognition (Acun and Yilmazer, 2023), and it can quickly extract acoustic and visual physical parameters at a specific site. In the future, the prediction of the soundscape perception evaluation results of the audio and video data collected on site through deep learning methods (Yang et al., 2025), combined with the recognition data of acoustic and visual physical parameters, will help to quickly explore the soundscape structural relationship in various types of open spaces.

## 5. Conclusion

This study used soundscape data collected from Beiling Park, Shenyang, China to establish a spatial mediation effect model to understand the relationship between sound source type saliencies, soundscape appropriateness, and acoustic comfort, the following conclusions were drawn:

1) Soundscape appropriateness mediates the effect of sound source type on acoustic comfort. Natural, traffic, and commercial sounds influence the perception of sound comfort by either enhancing or diminishing the appropriateness between the acoustic environment and the surrounding scene. By considering soundscape appropriateness, inner structural interconnections between soundscape variables were unveiled, offering empirical backing for the inherent interplay among sound, context, and soundscape quality.

2) The influence relationship between soundscape variables formed based on the mediating effect has the characteristics of spatial heterogeneity. This approach is able to identify a more accurate spatially resolved influence relationship by sound source type on soundscape appropriateness and acoustic comfort in localised spots and provide corresponding localised optimisation strategies.

## Authorship contribution

**Xinhao Yang:** Methodology, Formal analysis, Writing – original draft; **Guangyu Zhang:** Investigation, Methodology, Formal analysis; **Xiaodong Lu:** Writing - review & editing; **Yuan Zhang:** Conceptualisation, Writing - review & editing, Supervision, Funding acquisition; **Jian Kang:** Writing - review & editing, Project administration.

## Declaration of Competing Interest

The authors declare that they have no known competing financial interests or personal relationships that could have appeared to influence the work reported in this paper.

## Acknowledgements

Y.Z. acknowledges Educational Department of Liaoning Province Fundamental Research Project (No. JYTZD2023172). Y.Z. and J.K. acknowledge European Research Council (ERC) Advanced Grant (No. 740696)

## Appendix A. Supporting information

Supplementary data associated with this article can be found in the online version at https://doi.org/10.1016/j.jenvman.2024.123321.

# Supplementary

**for the manuscript entitled "Contribution of soundscape appropriateness to soundscape quality assessment in space: a mediating variable affecting acoustic comfort"**

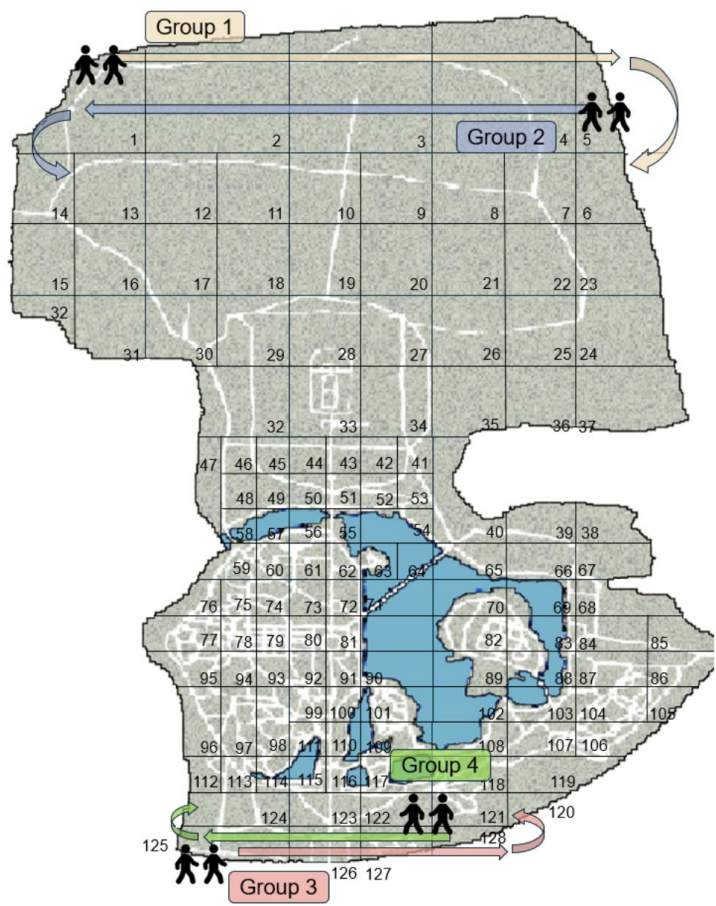

**Fig. A1** The grid layout of Beiling Park and the schematic representation of the soundwalk route.